\documentclass[twocolumn,aps,pre]{revtex4}
\usepackage{graphicx}
\usepackage{dcolumn}
\usepackage{amsmath}
\usepackage{latexsym}

\begin {document}

\title {Colored Percolation}
\author
{Sumanta Kundu and S. S. Manna}
\affiliation
{
\begin {tabular}{c}
Satyendra Nath Bose National Centre for Basic Sciences,
Block-JD, Sector-III, Salt Lake, Kolkata-700106, India
\end{tabular}
}
\begin{abstract}

      A model named `Colored Percolation' has been introduced with its infinite number of versions 
   in two dimensions. The sites of a regular lattice are randomly occupied with probability $p$ 
   and are then colored by one of the $n$ distinct colors using uniform probability $q = 1/n$. 
   Denoting different colors by the letters of the Roman alphabet, we have studied different 
   versions of the model like $AB, ABC, ABCD, ABCDE, ...$ etc. Here, only those lattice bonds 
   having two different colored atoms at the ends are defined as connected. The percolation 
   thresholds $p_c(n)$ asymptotically converges to its limiting value of $p_c$ as $1/n$. The model 
   has been generalized by introducing a preference towards a subset of colors when $m$ out of $n$ 
   colors are selected with probability $q/m$ each and rest of the colors are selected with 
   probability $(1 - q)/(n - m)$. It has been observed that $p_c(q,m)$ depends non-trivially on $q$ 
   and has a minimum at $q_{min} = m/n$. In another generalization the fractions of bonds between 
   similar and dissimilar colored atoms have been treated as independent parameters. Phase diagrams 
   in this parameter space have been drawn exhibiting percolating and non-percolating phases.
\end{abstract}

\maketitle

\section {Introduction}

      Over the last several decades, the phenomenon of percolation has been proved to be one of the most 
   investigated models in the topic of transport in random disordered systems \cite{Stauffer,Grimmett,Meester,Sahimi1, 
   Isichenko,Sahimi2}. Broadbent and Hammersley first introduced the model of percolation trying to 
   better understand the mechanism of fluid flow through a random porous medium \cite{Broadbent}, and 
   now it has become one of the simplest models of studying order-disorder phase transition \cite{Sornette}. 
   Due to its simplicity and plenty of applicability in a number of fields, the literature on this topic is 
   vast and expectedly a large number of variants of percolation models have been introduced to study the 
   critical behaviors of widely different systems \cite{Sahimi2, Araujo,Saberi,Lee,Santi,EP,Manna,DP,Morone}. 
   In particular, the percolation theory has been successfully applied to the well known sol-gel transition 
   \cite{Coniglio}, transitions in conductor-insulator mixtures using the random resistor networks 
   \cite{Arcangelis,Batrouni}, propagation of fires in the forests \cite{Stauffer,Beer}, spreading of 
   infectious diseases in the form of epidemics \cite{Newman,Tome}, etc.

      In the ordinary percolation, the sites of a regular lattice are occupied randomly and 
   independently with probability $p$ or kept vacant with probability ($1-p$). Any two adjacent
   occupied sites are considered as connected. A group of such occupied sites interconnected through 
   their neighboring connections forms a cluster, the properties of which depend on $p$. At any 
   arbitrary value of $p$, there are several clusters of different shapes and sizes. The size of 
   the largest cluster increases monotonically as $p$ is increased and right above a critical value of
   $p = p_c$, known as the percolation threshold, the largest cluster includes sites on the opposite 
   sides of the lattice and thus for the first time a global connectivity is established. Therefore, 
   $p_c$ marks the transition point, between the globally connected and disconnected phases, characterized 
   by the divergence of the correlation length $\xi(p)$ as $p \to p_c$. It is well known that the ordinary 
   percolation undergoes a continuous phase transition at $p = p_c$ and the set of critical exponents 
   defined at and around $p_c$ characterizes the universality class of the transition \cite{Stauffer}. The 
   best known value of $p_c$(sq) for the site percolation on the square lattice is 0.59274605079210(2) 
   \cite{Jacobsen} and 1/2 for the bond percolation \cite{Ziff-Wiki}. 

\begin{figure*}[t]
\begin {tabular}{cccc}
\includegraphics[width=4.3cm]{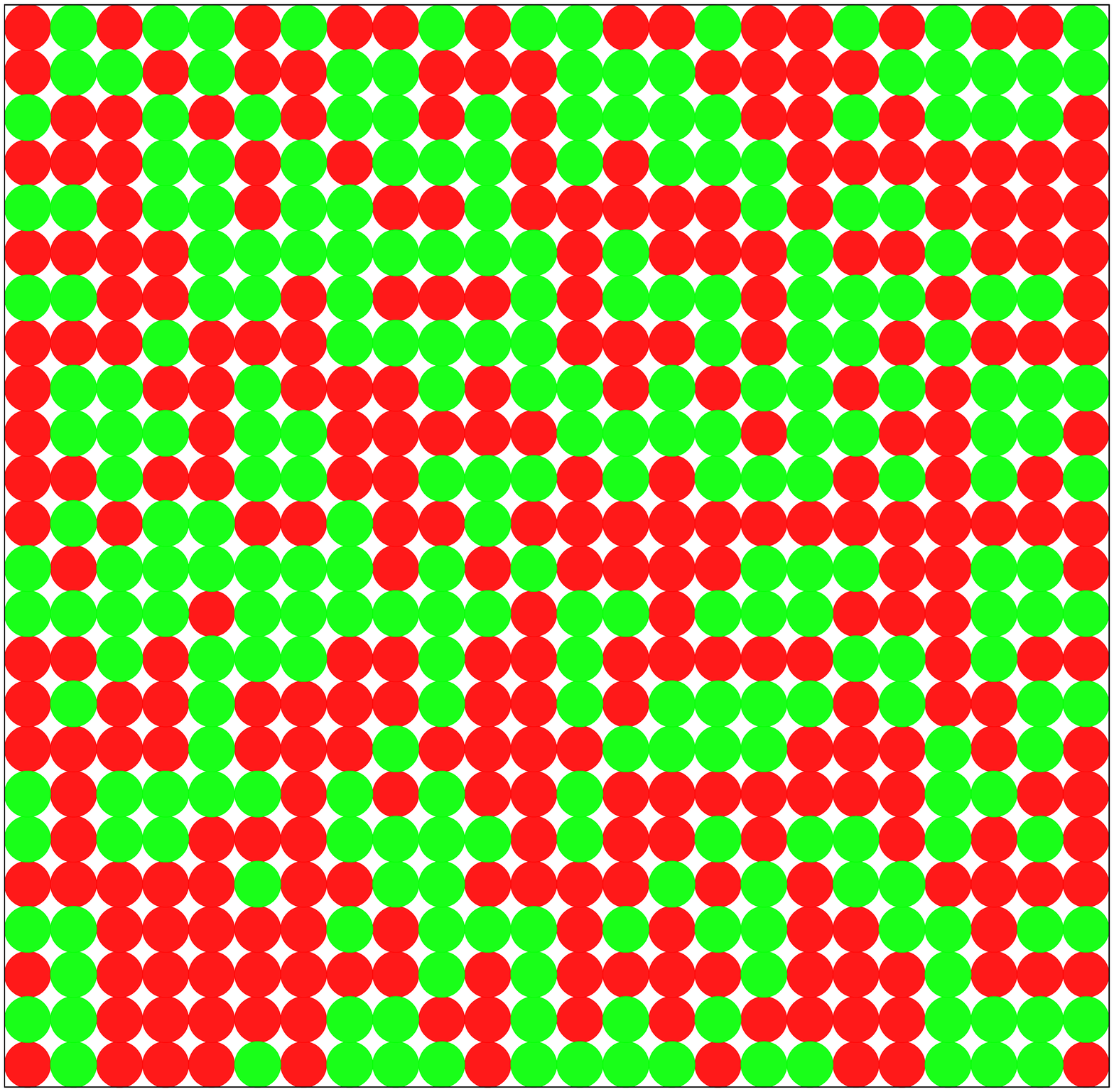} & \includegraphics[width=4.3cm]{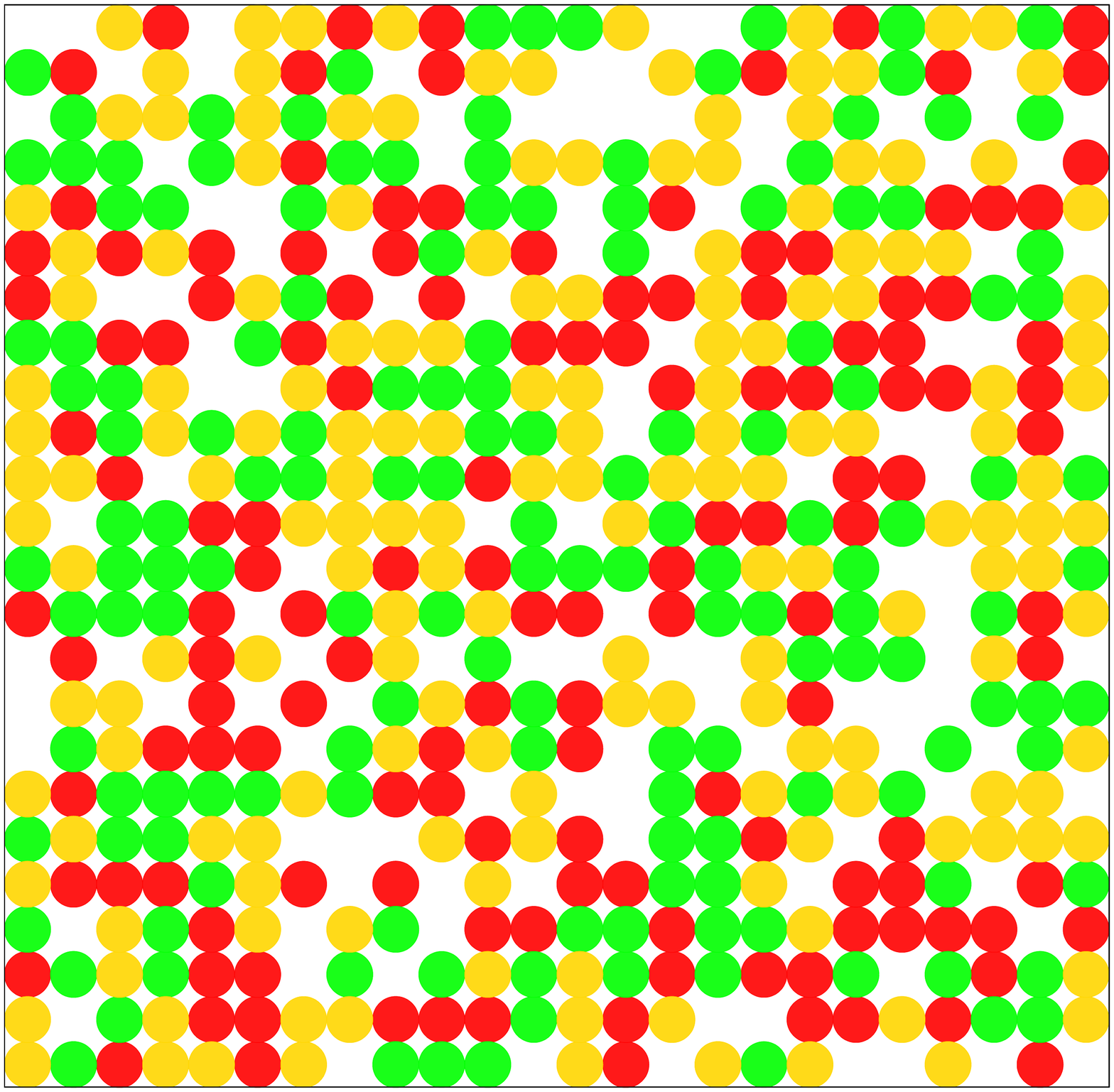} &
\includegraphics[width=4.3cm]{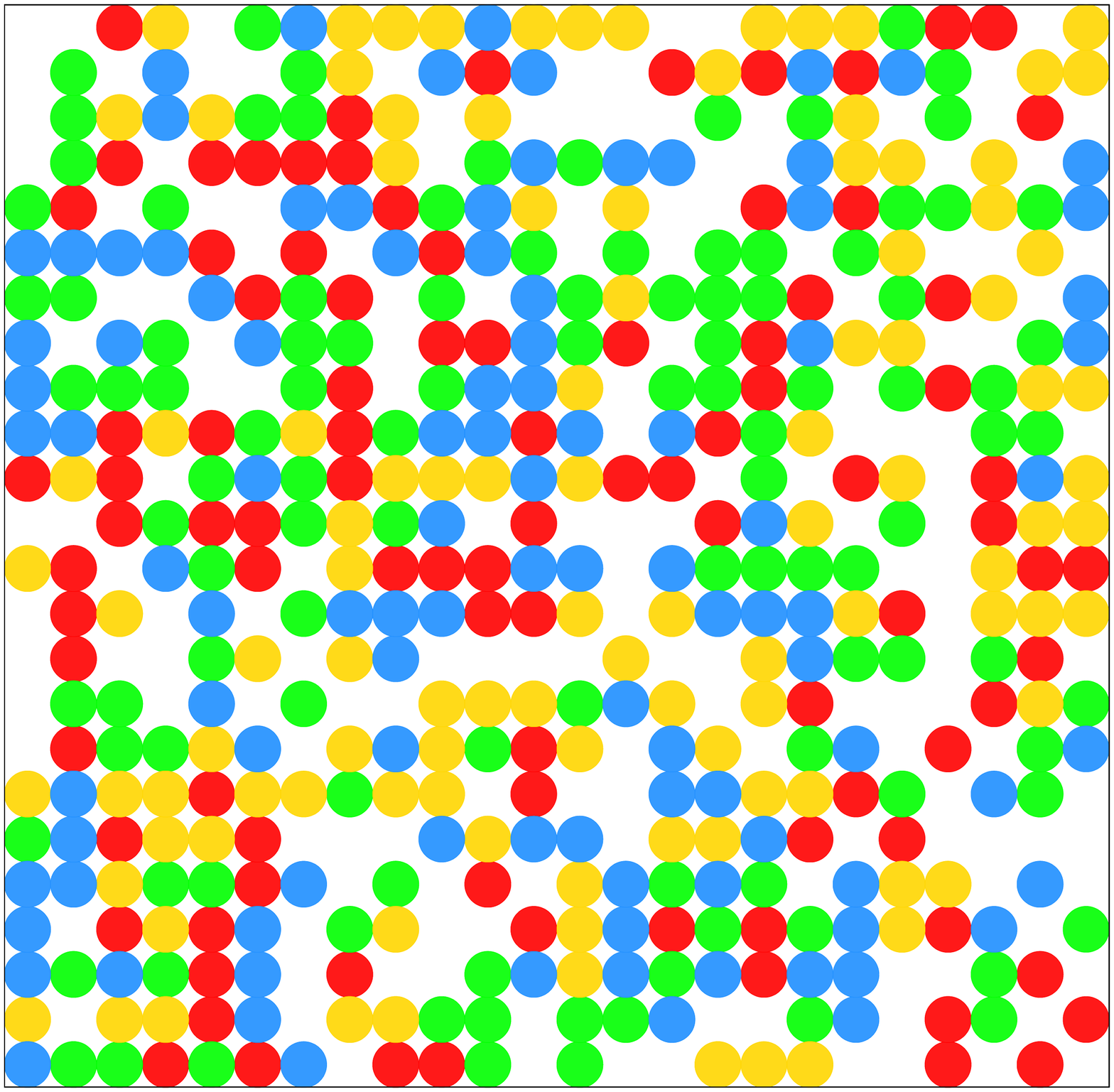} & \includegraphics[width=4.3cm]{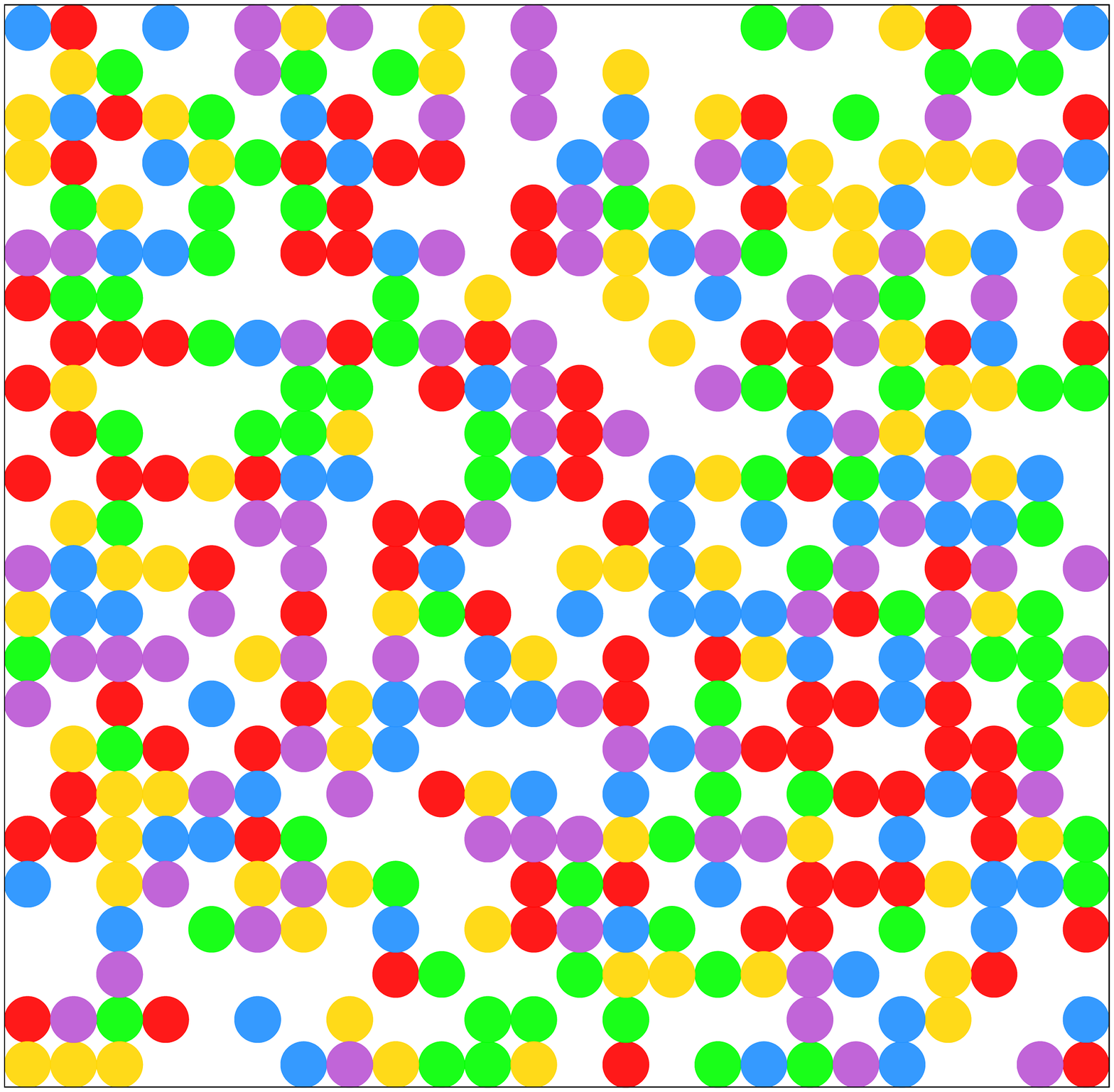} \\
\includegraphics[width=4.3cm]{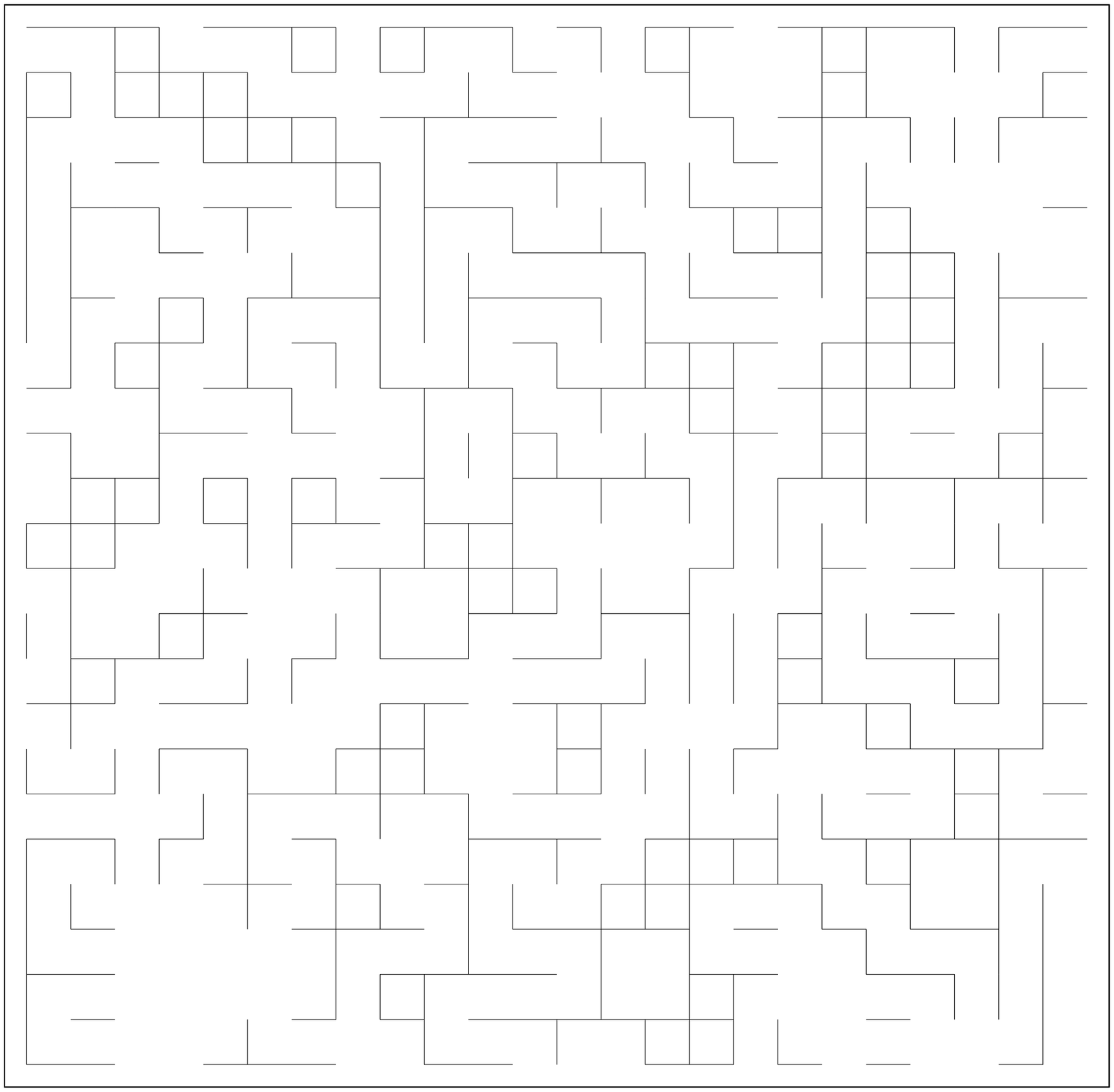} & \includegraphics[width=4.3cm]{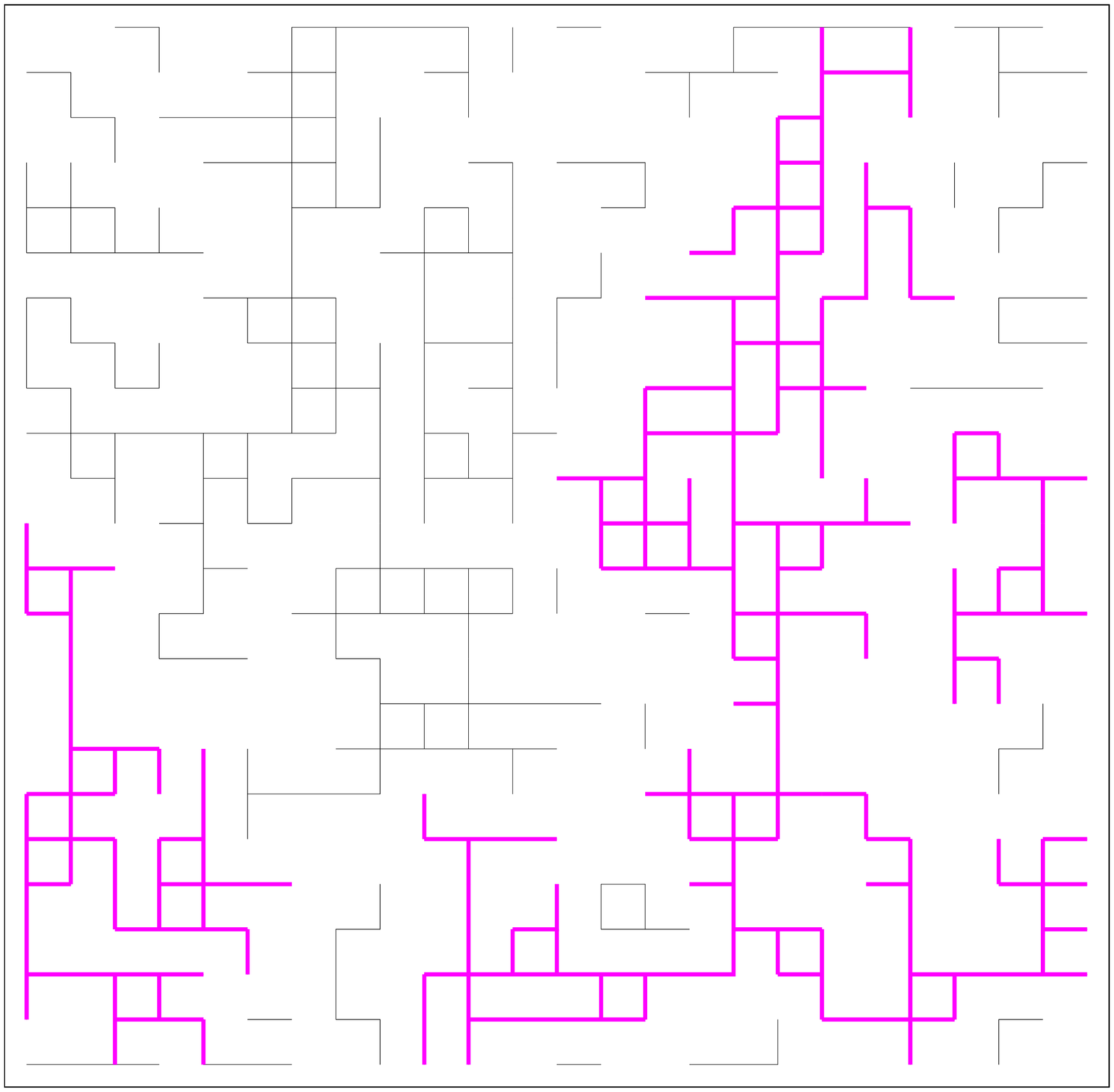} &
\includegraphics[width=4.3cm]{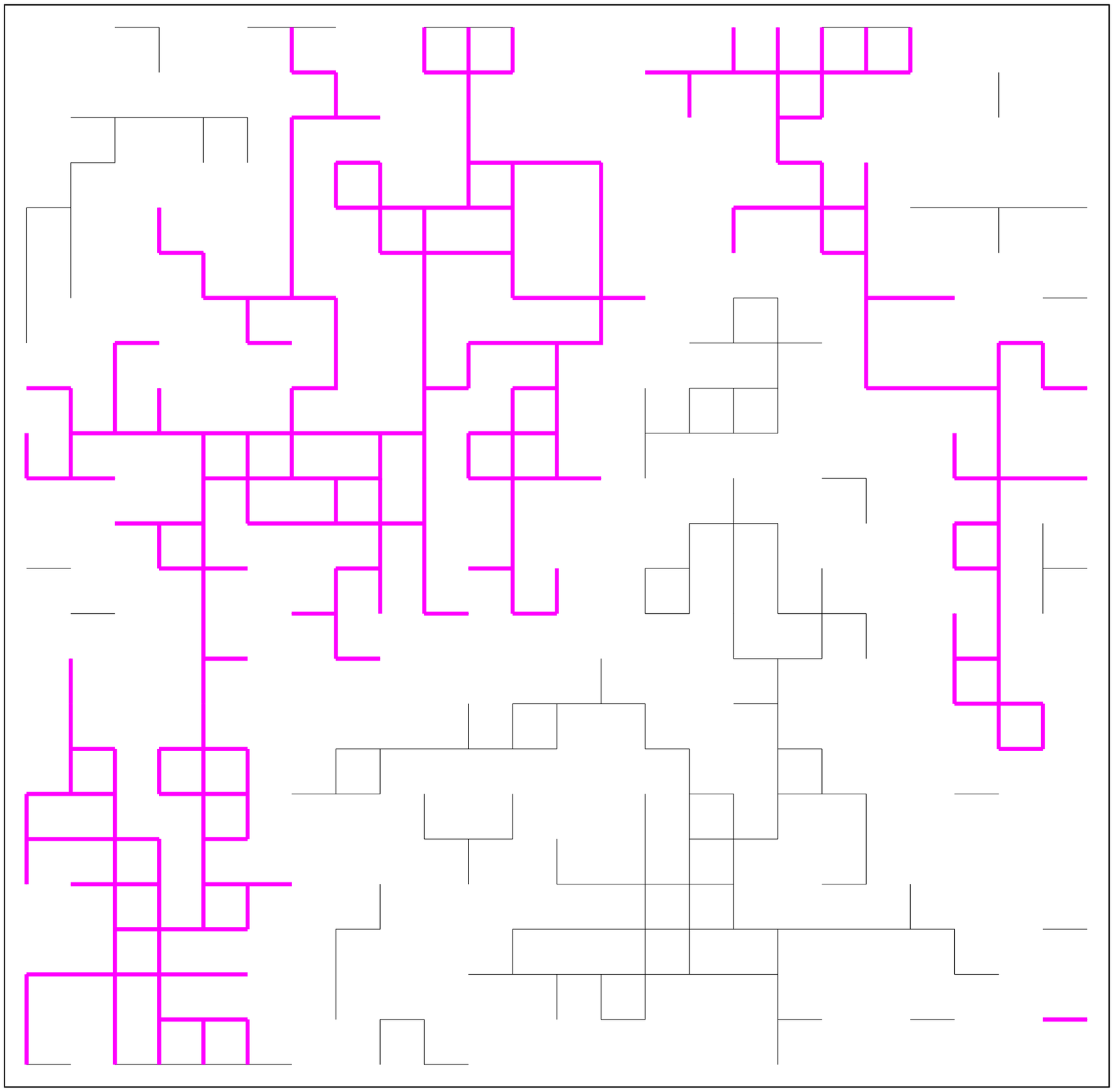} & \includegraphics[width=4.3cm]{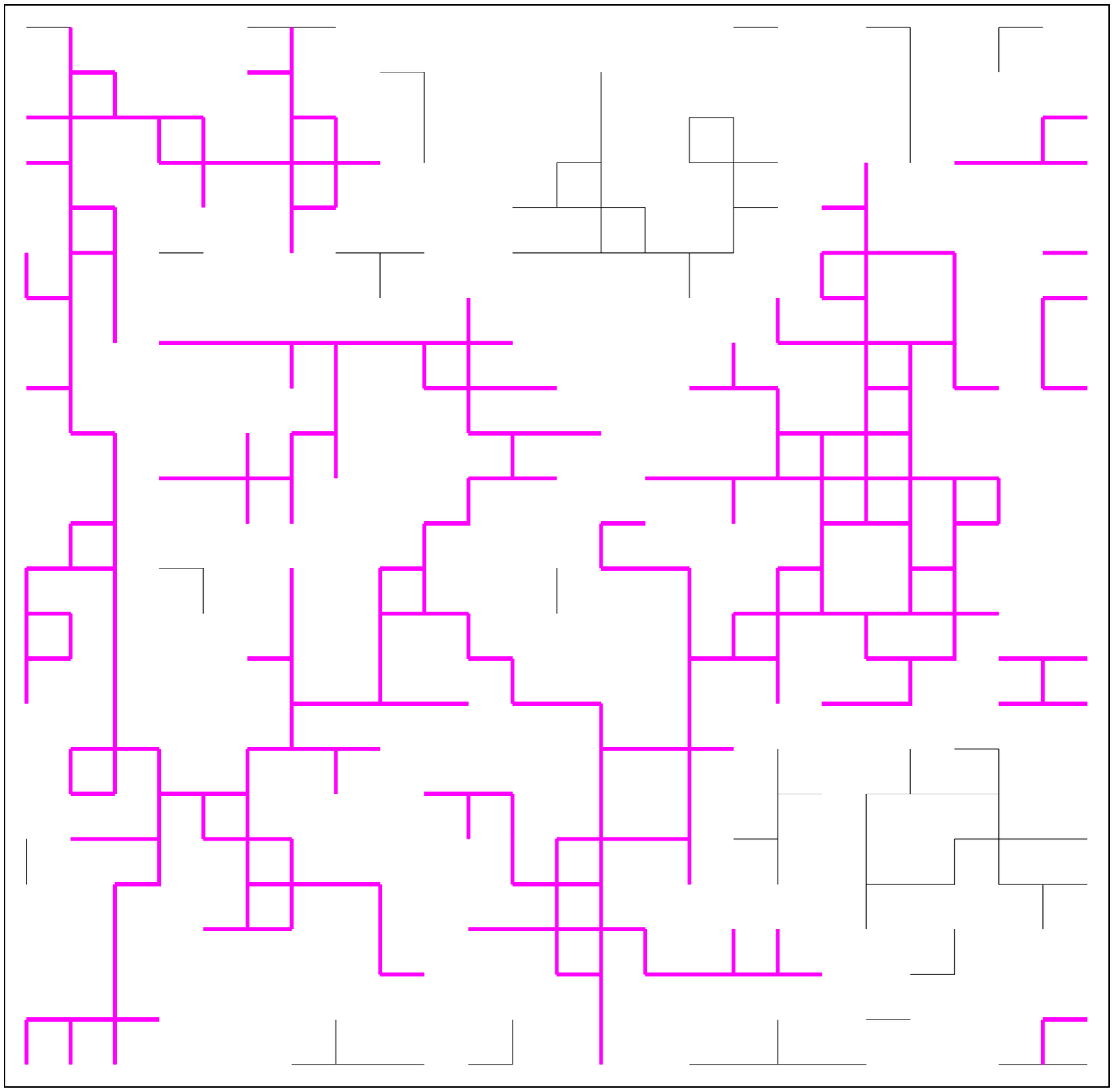}
\end {tabular}
\caption{
   Typical configurations of AB, ABC, ABCD, and ABCDE percolation (from left to right) at their 
   critical points have been shown on a $24 \times 24$ square lattice with periodic boundary 
   conditions along the horizontal direction. The colors used are: red ($A$), green ($B$), 
   yellow ($C$), blue ($D$) and orchid ($E$) and are distributed with uniform probabilities. 
   The corresponding bond configurations with the spanning clusters painted in magenta have been 
   shown in the lower panel. It may be noted that for $AB$ percolation, there is no spanning cluster even 
   when all the sites are occupied. 
}
\label {FIG01}
\end{figure*}

      Inspired by the phenomena of anti-ferromagnetism, gelation, spreading of infection from the infected 
   cells to normal cells, Mai and Halley introduced the $AB$ percolation model \cite{Grimmett,Mai,Barma,Wierman}.
   The model of $AB$ percolation is illustrated in the following way. Initially, all sites of a lattice 
   are occupied with $B$ atoms. Then, random sites are selected one by one and the $B$ atoms at these sites 
   are replaced by the $A$ atoms. At any arbitrary intermediate stage the fraction of $A$ atoms is denoted 
   by $r$. According to this model, bonds which are having both $A$ and $B$ atoms at their opposite ends 
   are marked connected. For a given value of $r$, the probability that any given edge has a bond is 2$r(1-r)$,
   which has its maximum at $r = 1/2$ and decreases monotonically on both sides of this point. Consequently, 
   the average size of the largest cluster gradually grows till $r$ = 1/2. The entire scenario is symmetric 
   about $r$ = 1/2. At $r = 1 - r_c$, the size of the largest cluster drops sharply, and finally it vanishes 
   at $r = 1$. Therefore, for $r_c \leq r \leq 1 - r_c$, the system is percolating. However, the existence 
   of a global connectivity through the alternating $A$ and $B$ atoms and therefore the existence of $r_c$, 
   crucially depends on the geometry of the underlying lattice \cite{Wierman,Appel}. For example, spanning 
   $AB$ cluster does not exist on the square lattice \cite{Appel,Wu}, but it exists on the triangular lattice 
   \cite{Wierman2}. Although, it was first concluded that the universality class of $AB$ percolation in two 
   dimensions is different from the ordinary percolation \cite{Mai}, later it has been argued that it belongs 
   to the same universality class as the ordinary percolation \cite{Sevsek,Nakanishi,Wilkinson}. Further, random 
   occupation of lattice sites by more than two distinct atoms was studied through the model of polychromatic 
   percolation \cite{Zallen,Halley}. 

      In this paper, we consider a percolation model, where the sites of a regular lattice are occupied 
   with probability $p$ similar to the ordinary site percolation, and then at every occupied site one 
   of the $n$ different colored atoms is assigned with a given probability $q$. A bond between a pair of 
   neighboring occupied sites is declared as connected if the atoms are of different colors. We refer 
   this model as the colored percolation. We study the critical properties of this model for both the 
   square and triangular lattices. 

      The paper is organized as follows. We start by describing the model of colored percolation in Sec. II, 
   where we consider every color to be equally likely. In Sec. III, we generalize the model by introducing 
   a preference towards the selection of a subset of colors and its simulation results. Percolation
   transition using similarly colored bonds in addition to the dissimilar bonds has been described in Sec. IV. 
   Percolation transition mixing the fractions of similarly and dissimilarly colored bonds are reported in Sec.
   V. The critical properties of the model are presented in Sec. VI. Finally, we summarize in Sec. VII.

\section {Model}

      The sites of a $L \times L$ regular and initially empty lattice are occupied randomly by atoms one by one. 
   At any arbitrary stage, the density of atoms is denoted by $p$. After occupying a given site, the corresponding 
   atom is colored by selecting one of the $n$ colors with probability $q = 1/n$. The letters of the Roman alphabet
   are used to denote the different colors. The bonds, which have two distinct colored atoms at their opposite ends, 
   are declared as connected. Therefore, the other bonds having same colored atoms like $AA$, $BB$ etc., are not 
   connected. Gradually, the number of connected bonds in the system increases with increasing the value of $p$. 

      For $n$ = 2, every selected site is occupied either by a $A$ atom or by a $B$ atom with probability $q = 1/2$. 
   Like the model of $AB$ percolation, the size of the largest cluster never assumes a macroscopic size on the square 
   lattice and therefore, a percolation transition is absent here. On the other hand, our simulation results indicate
   that on the triangular lattice the percolation transition occurs at $p_c \approx 0.729$.
   
      For $n = 3$, every selected site is occupied by one of the three atoms $A$, $B$, and $C$ with probabilities $q = 1/3$. 
   Only the $AB$, $BC$, and $CA$ bonds are defined to be connected. In this case, there exists a  percolation 
   threshold $p_c \approx 0.807$ on the square lattice and $\approx 0.630$ on the triangular lattice. We refer 
   this model as $ABC$ percolation.
   
      Such an extension of the model can be continued with four colored atoms, where $A$, $B$, $C$, and $D$ 
   atoms are distributed with probabilities $q = 1/4$. The critical densities of the occupied sites are $p_c 
   \approx 0.734$ on the square lattice and $\approx 0.591$ on the triangular lattice estimated using the method 
   in \cite {Kundu}. This model is referred as the $ABCD$ percolation.

      We systematically increase the number of distinct colored atoms to define further the $ABCDE, ABCDEF, ABCDEFG$ 
   etc., colored percolation models. In brief, we have been able to define an infinite set of percolation models by 
   defining connectivity through the bonds between dissimilar atoms. Figure \ref{FIG01} shows the images of the typical 
   percolation configurations on the square lattice for four different values of $n$. 

\subsection {The Order Parameter and the Percolation Threshold}

\begin{figure}[t]
\begin {center}
\includegraphics[width=6.0cm]{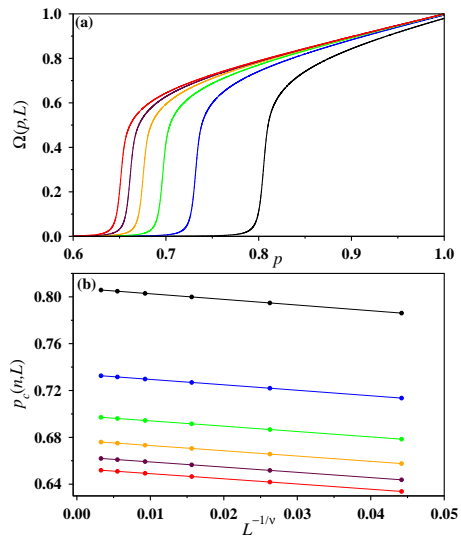}
\end {center}
\caption{
   For $n$ = 3 (black), 4 (blue), 5 (green), 6 (orange), 7 (maroon), and 8 (red): (a) the order parameter 
   $\Omega(p,L)$, has been plotted against the site occupation probability $p$ for the square lattice of 
   size $L = 1024$ ($n$ increases from right to left); (b) plot of the percolation thresholds $p_c(n,L)$ against $L^{-1/\nu}$ using $\nu = 4/3$
   ($n$ increases from top to bottom).
   By extrapolating as $L \to \infty$, we obtain the asymptotic values of the percolation threshold $p_c(n)$.
}
\label {FIG02}
\end{figure}

      The size of the largest cluster for a particular value of $p$ and for the system of size $L$ is denoted 
   by $s_{max}(p,L)$ and the average fractional size of the largest cluster is defined as the order parameter 
   $\Omega(p,L) = \langle s_{max}(p,L) \rangle / L^2$. The variation of $\Omega(p,L)$ with $p$ has been shown 
   in Fig. \ref{FIG02}(a) for six different values of $n$ on the square lattice. The sharp rise in the order 
   parameter curve shifts towards smaller values of $p$ with increasing the value of $n$. 
   
      As $p$ is increased, the size of the largest cluster increases monotonically by merging with the other 
   clusters, whereas the variation of the second largest cluster is not monotonic. In a typical run $\alpha$, 
   the second largest cluster may merge several times with the largest cluster and thereby causes multiple 
   jumps in the size of the largest cluster. At a specific value of $p$, the maximum of the second largest 
   cluster merges with the largest cluster that results the maximal jump in the size of the largest cluster. 
   This particular value of $p$ is defined as the percolation threshold $p_c^\alpha$ for the run $\alpha$ 
   \cite{Kundu,Margolina}. For a fixed value of $n$, this calculation is repeated over a large number of 
   independent runs $\alpha$ and the $p_c^\alpha$ values are averaged to obtain $p_c(n,L) = \langle p_c^\alpha \rangle$ 
   for the system size $L$. In our simulation, periodic boundary conditions are imposed along both the vertical 
   and horizontal directions. The value of $s_{max}(p,L)$ is evaluated using the algorithm given in reference
   \cite{Newman2} over the entire range of $p$.
     
      For a given value of $n$, the $p_c(n,L)$ values are extrapolated using Eqn. \ref{EQN01},
\begin{equation}
   p_c(n,L) = p_c(n) - AL^{-1/\nu}
\label {EQN01}
\end{equation}
   to obtain the asymptotic value of the percolation threshold $p_c(n)$ for $L \to \infty$, where $\nu$ is known as 
   the correlation length exponent. Using $1/\nu$ as a free parameter we varied it's trial values at the interval of 0.001
   and found by the least square fitting method that the best values for all $n$ differ from 3/4 by at most 0.005. Therefore in the
   rest of our calculation we have used $\nu = 4/3$, the exact value of the exponent in two dimensions 
   \cite{Stauffer,Eschbach}. In Fig. \ref{FIG02}(b), we plot $p_c(n,L)$ against $L^{-1/\nu}$ in 
   a linear scale for six different values of $n$. The data points for all six values of $n$ fits excellently 
   to a straight line. By extrapolating the straight lines as $L \to \infty$ and measuring the $y$-intercept 
   we estimate the asymptotic values of $p_c(n)$. The values of $p_c(n)$ for first few values of $n$ are 
   listed in Table I for square and triangular lattices. It is to be noted that, for each value of $n$, 
   the $p_c(n,L)$ values are calculated numerically using $L$ = 64, 128, 256, 512, 1024, and 2048.

\begin{table}[b]
\begin {tabular}{ccc} 
\hline 
\multicolumn{3}{c}{\hspace*{0.3cm}$p_c(n)$} \\ 
\cline{2-3}
 \hspace*{0.3cm}   n  \hspace*{0.3cm} & Square   \hspace*{0.6cm} & Triangular        \\ \hline \hline 
   2  &                     & 0.72890(4)  \\
   3  & 0.80745(5)          & 0.63005(4)  \\
   4  & 0.73415(4)          & 0.59092(3)  \\
   5  & 0.69864(7)          & 0.56991(5)  \\
   6  & 0.67751(5)          & 0.55679(5)  \\
   7  & 0.66345(5)          & 0.54782(3)  \\
   8  & 0.65342(8)          & 0.54130(3)  \\
   9  & 0.64588(5)          & 0.53634(2)  \\
  10  & 0.64002(4)          & 0.53245(3)  \\
  11  & 0.63532(5)          & 0.52931(2)  \\
  12  & 0.63147(4)          & 0.52672(2)  \\ \hline \hline
\end {tabular}
\caption{Numerical estimates of the asymptotic values of the percolation threshold $p_c(n)$ for $n$ different 
   colored atoms occur with probability $1/n$ for the square and triangular lattice geometries.
}
\end {table}

      As the number of different colors increases, there are more and more connected bonds in the system 
   which helps the system to percolate at smaller densities. The probability that a bond would be a 
   connected one is given by $p_b = 1 - 1/n$ and therefore, for very large value of $n$ the entire scenario 
   is exactly same as the ordinary site percolation. Expectedly, $p_c(n)$ approaches towards $p_c = p_c(\infty)$, 
   the ordinary site percolation threshold on the corresponding lattice. 

\begin{figure}[t]
\begin {center}
\includegraphics[width=6.0cm]{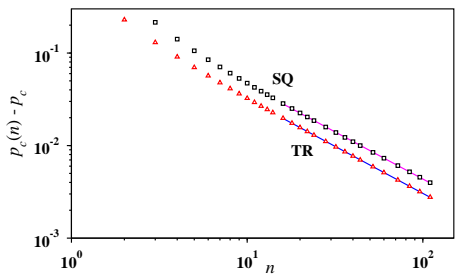}
\end {center}
\caption{Plot of $p_c(n) - p_c$ against $n$ (upto $n = 110$) on a $\log - \log$ scale, using $p_c = p_c$(sq) and 1/2 for the square and 
   triangular lattices respectively. The slope of the curves in the fitted regime have been found to 
   be 1.020 and 1.017 respectively.}
\label {FIG03}
\end{figure}

      To investigate how the asymptotic values of the percolation threshold $p_c(n)$, approach to the value 
   $p_c$ as $n \to \infty$, we first calculate $p_c(n)$ for different values of $n$ upto $n$ = 110. Then, we 
   plot the deviation $p_c(n) - p_c$ against $n$ on a double logarithmic scale for the square and triangular 
   lattices (Fig. \ref{FIG03}). Although both the curves have curvatures at their initial regimes, for 
   large values of $n$ they are quite straight. The slopes of the curves within a window ranges between 
   $n$ = 16 and 110 measured as 1.020 and 1.017 for the square and triangular lattices respectively. We observed 
   that these slopes approach gradually to a value unity as we shift the window to higher values of $n$. Thus, 
   we conjecture that $p_c(n) - p_c \sim n^{-1}$ for both the lattices.
   
\section {Preferential Colored Percolation}

      A straightforward generalization of this model can be achieved by introducing a preference towards 
   the probability of selection of different colored atoms. In the simplest case, let us consider the 
   atoms of color $C$ to be preferentially selected whereas all other colored atoms are on the same footing.
   More specifically, we denote the probability of selection of the $C$ atoms by $q$ and for all other atoms it 
   is $(1-q)/(n-1)$. As before, only the bonds between dissimilar atoms are defined to be connected. 

\begin{figure}[t]
\begin {center}
\includegraphics[width=6.0cm]{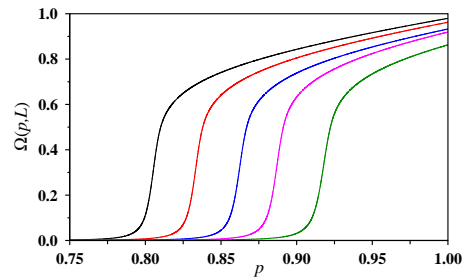}
\end {center}
\caption{Plot of the order parameter $\Omega (p,L)$ against occupation probability $p$ for the preferential
   $ABC$ percolation on square lattice of size $L$ = 1024. The values of $q$ are 1/3 
   (black), 0.20 (red), 0.53 (blue), 0.12 (magenta), and 0.60 (green) arranged from left to right.
}
\label {FIG04}
\end{figure}

      For a given value of $q$, if $\text{Prob}_{ii}$ is the probability that two atoms at the end sites of a 
   bond are of same color $i$, then the probability that a particular bond would be a connected one 
   is given by,
\begin{equation} 
\text{Prob}_b(q) = 1 - \sum\limits_{i=1}^n \text{Prob}_{ii} = 1 - q^2 - (1-q)^2 / (n-1).
\label{EQN02}
\end{equation} 
   The above expression describes that with increasing the value of $q$ from $0$, the $\text{Prob}_b(q)$ first 
   increases, reaches its maximum at $q = q_{min}$ and finally decreases beyond this point. The condition 
   $d\text{Prob}_b(q)/dq$ = 0 at $q = q_{min}$ yields the value of $q_{min} = 1/n$. This property of $\text{Prob}_b(q)$ 
   should be reflected in the percolation properties of the system also. Consequently, the percolation threshold 
   $p_c(q)$ must decrease with $q$, till it reaches $q_{min}$ and then increase for $q > q_{min}$.
   
\begin{figure}[t]
\begin {center}
\includegraphics[width=6.0cm]{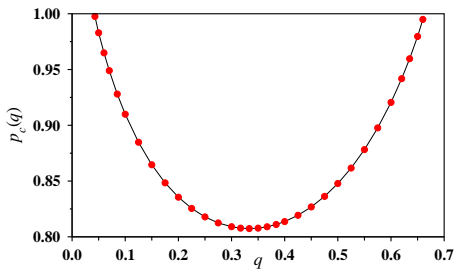}
\end {center}
\caption{For the preferential $ABC$ percolation on the square lattice, percolation threshold $p_c(q)$ is plotted 
   against the parameter $q$. The minimum of $p_c(q)$ occurs at $q=0.333$ which is in agreement with its 
   estimate of 1/3, using Eqn. (\ref{EQN02}).
}
\label {FIG05}
\end{figure}

     For $n = 3$, the variations of the order parameter for different values of $q$ have been shown in Fig. \ref{FIG04}. 
   For $q = 0$, one gets back the $n = 2$ unpreferred colored percolation, and therefore, for the square lattice even a 
   fully occupied lattice does not percolate.  Further, on increasing the value of $q$ the $BC$ and $CA$ bonds are 
   created which eventually also contribute to the global connectivity. Tuning the value of $q$, it has been observed 
   that there exists a threshold value of $q = q_1$ when the global connectivity first appears, i.e., $p_c(q_1) = 1$. 
   If the probability $q$ is increased further, the percolation threshold $p_c(q)$ gradually decreases and reaches to 
   its minimum value $p_c(q_{min}) \approx 0.807$ at $q_{min} \approx 0.333$. On increasing $q$ even further, $p_c(q)$ 
   increases and reaches the value of unity again at $q = q_2$. The global connectivity is lost beyond this point. For 
   each value of $q$, first the percolation threshold $p_c(q,L)$ for the system size $L$ is estimated for $L$ = 256, 512 
   and 1024 and then they are extrapolated as $L \to \infty$ using Eqn. (\ref{EQN01}) to obtain $p_c(q)$.  The variation 
   of $p_c(q)$ for the entire range of $q$ has been shown in Fig. \ref{FIG05}.

      For the triangular lattice, the curve retains its shape but in this case, even for $q$ = 0, there exists 
   a percolation threshold. Since, for $q$ = 0 the other two atoms are occupied with probability 1/2, the value 
   of percolation threshold is expected to be $\approx$ 0.729. Similarly, for $q$ = 0 and $n$ = 4 on the square 
   lattice this model is identical to the $ABC$ percolation and therefore, $p_c(0) \approx 0.807$.

\begin{figure}[b]
\begin {center}
\includegraphics[width=6.0cm]{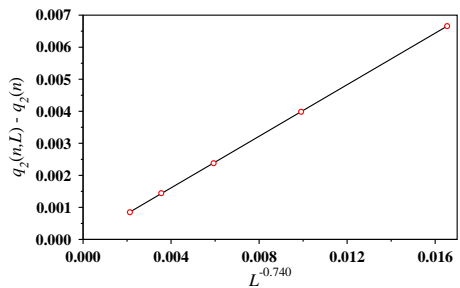}
\end {center}
\caption{For $n$ =3 and for square lattice, plot of $q_2(n,L) - q_2(n)$ against $L^{-0.740}$ with $q_2(n)$ = 0.6639 
   exhibits an excellent straight line which passes very close to the origin.
   The value of $q_2(n,L)$ is calculated for $L$ = 256, 512, 1024, 2048 
   and 4096.
}
\label {FIG06}
\end{figure}

      Numerically, the values of $q_1$ and $q_2$ are determined using the bisection method in the following way. 
   To estimate $q_2$, we select a pair of values $q^c$ and $q^d$ for $q$ so that the system is globally connected 
   and disconnected respectively for $p = 1$. We have applied the periodic boundary condition along one direction 
   and tested for the global connectivity along its transverse direction using the Burning algorithm \cite{Herrmann} 
   for $q = (q^c+q^d)/2$. If the system is globally connected then $q^c$ is replaced by $q$, otherwise $q^d$ is 
   replaced by $q$. This procedure is repeated until $q^d - q^c < 10^{-7}$, when $(q^c + q^d)/2$ defines $q_2$ for 
   a particular run. Averaging over a large number of independent runs $q_2(n,L)$ for specific values of $n$ and $L$
   has been estimated. Similar procedure has been followed for $q_1(n,L)$. The entire procedure is then repeated for 
   different values of $L$ and extrapolations to $L \to \infty$ using Eqn. (\ref{EQN01}) in this case as well, we
   obtain $q_1(n)$ and $q_2(n)$. For $n = 3$, the best linear fit has been exhibited in Fig. \ref{FIG06} by plotting 
   $q_2(3,L) - q_2(3)$ against $L^{-0.740}$, using $q_2(3) = 0.6639(5)$ on the square lattice. Similarly, $q_1(3) = 
   0.0414(5)$ has been estimated.

      Therefore, a non-trivial value of $q_1(n)$ exists for $n = 2$ on the square lattice. It does not exist for
   all other values of $n$ on the square lattice and for all values of $n$ on the triangular lattice.
   On the other hand, $q_2(n)$ exists for both the lattices and for all values of $n$. 
   
      The density of connected bonds $\text{Prob}_b(q)$ corresponding to the point $q = q_2(n)$ represents a threshold value 
   in the correlated bond percolation scenario. Beyond this point, the density of connected bonds is no longer sufficient to 
   establish a global connectivity. Neglecting the local correlations and equating $\text{Prob}_b(q_2)$ to $p_c^b$, the 
   random bond percolation threshold of the respective lattices, we arrive at an expression of $q_2(n)$ using Eqn. (\ref{EQN02}),
\begin{equation} 
q_2(n) = \Big[1 + \big[1 + n[(1 - p_c^b)(n-1)-1]\big]^{1/2}\Big]\big/n.
\label{EQN03}
\end{equation} 
   Numerically estimated values of $q_2(n)$ for different values of $n$ using bisection method, along with the 
   values obtained from Eqn.  (\ref{EQN03}) using $p_c^b$ = 1/2, are summarized in Table II for the square lattice. 
   It is observed that the values are close to each other and differ only due to the existence of short range 
   correlations in the system.

\begin{table}[b]
\begin {tabular}{lcccccc} \\ \hline \vspace*{0.1cm}
   n          & 3      & 4      & 5      & 6      & 7      & 8       \\ \hline \hline
Numerical     & 0.6639 & 0.6849 & 0.6927 & 0.6969 & 0.6995 & 0.7013  \\
Analytical    & 2/3    & 0.6830 & 0.6899 & 0.6937 & 0.6961 & 0.6978  \\ \hline \hline
\end {tabular}
\caption{The comparison of $q_2(n)$, evaluated using Eqn. (\ref{EQN03}) with its numerical estimates for different 
   values of $n$ for square lattice. For each values of $n$, numerically $q_2(n,L)$ is calculated for $L$ = 256, 512, 
   1024, 2048, and 4096 and on extrapolation to $L \to \infty$ we obtained $q_2(n)$. Each of the reported value has 
   an error bar of 5 in the last digit.
}
\end {table}
\begin{figure}[t]
\begin {center}
\includegraphics[width=6.0cm]{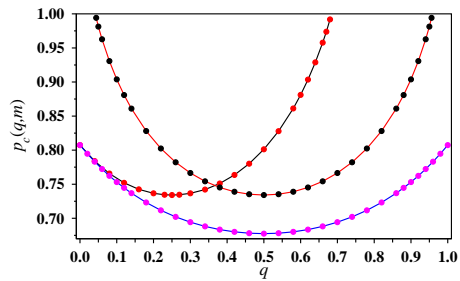}
\end {center}
\caption{The variation of the percolation threshold $p_c(q,m)$ with $q$ for the generalized version of the preferential colored
   percolation model is shown for $m$ = 1, $n$ = 4 (black); $m$ = 2, $n$ = 4 (red); and $m$ = 3, $n$ = 6 (blue) for 
   the square lattice. The curves are arranged from top to bottom along the line $q$ = 0.50.
}
\label {FIG07}
\end{figure}

      A more general version of the preferential colored percolation model is the situation when $m$ distinct colored atoms are
   equally probable and the remaining $(n-m)$ colors are also equally probable but occur with different probabilities.
   Such a generalization can be obtained by assigning an atom of one of the $m$ 
   colors with probability $q/m$ and rest of the $(n-m)$ colors with probability $(1-q)/(n-m)$ at the time of 
   occupying a vacant site. The probability for an arbitrary bond to be connected is given by,
\begin{equation} 
\text{Prob}_b(q,m) = 1 - q^2 / m - (1-q)^2 / (n-m).
\label{EQN04}
\end{equation} 
   Evidently, $\text{Prob}_b(q,m)$ is maximum at $q_{min} = m/n$ and it decreases on both sides of this point. The expression of 
   $\text{Prob}_b(q,m)$ remains unaltered if the value of $q$ is changed from $q$ to $(1-q)$ at the same time $m$ is changed
   from $m$ to $(n-m)$, i.e., $\text{Prob}_b(q,m) = \text{Prob}_b(1-q, n-m)$. Immediately, it implies that the curve is 
   symmetric about 
   $q = 1/2$ only when $m = n/2$. The percolation threshold $p_c(q,m)$ for specific values of $n$ and $m$ is expected 
   to exhibit such properties appropriately.

      Again, after extrapolation to the large $L$ limit using Eqn. (\ref{EQN01}) with $\nu = 4/3$ we obtain $p_c(q,m)$.
   In Fig. \ref{FIG07}, the asymptotic values of the percolation threshold $p_c(q,m)$ have been plotted against $q$ for 
   three pairs of values of $m$ and $n$ for the square lattice. It is observed that all three curves have their own 
   minimum which occur at $q_{min}$ = 0.25, 0.50 and 0.50 for $m$ = 1, $n$ = 4; $m$ = 2, $n$ = 4; and $m$ = 3, $n$ = 6 
   respectively. Clearly, the $q_{min}$ values match excellently with our analytically estimated value of $q_{min} = m/n$. 
   As expected, the curves corresponding to $m = n/2$ are symmetric about the point $q$ = 1/2.

\begin{figure}[t]
\begin {center}
\includegraphics[width=6.0cm]{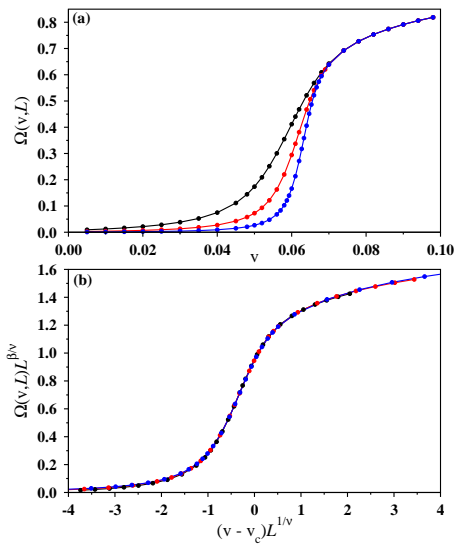}
\end {center}
\caption{(a) For $L$ = 256 (black), 512 (red) and 1024 (blue) (arranged from left to right), the order parameter $\Omega(v,L)$ has been plotted with the bond 
   occupation probability $v$ between $AA$ or $BB$ atoms at $p = 1$, for $n = 2$ with $q = 1/2$ colored percolation.
   (b) By suitably scale the abscissa and ordinate when the same data as in (a) is replotted, a nice data collapse 
   is observed using $v_c$ = 0.0651, $\beta/\nu = 0.101(5)$ and $1/\nu = 0.745(5)$.
}
\label {FIG08}
\end{figure}

\section {Percolation using additional similar bonds}

      Let us recall that for $n = 2$ case on the square lattice and for any arbitrary value of the site occupation 
   probability $p$, the density of $AB$ bonds is maximum for $q = 1/2$. In spite of that no percolation transition 
   is observed on the square lattice since the largest cluster of $AB$ bonds is found to be minuscule even when 
   $p = 1$. In other words, even the maximum number of connected bonds are not sufficient to establish a global 
   connectivity across the system \cite{Appel}.

      In this section we study a new variant of our colored percolation model, where in addition to the $AB$ bonds, we 
   allow also a fraction $v$ of similarly colored bonds (like $AA$ and $BB$) to be connected. Therefore, for $v = 1$, 
   the problem reduces to the ordinary site percolation with the percolation threshold at $p_c$(sq). This suggests
   that for all values of $p_c$(sq) $ \leq p \leq 1$ there should be a critical value of $v = v_c(p)$ for the fraction 
   of bonds between similarly colored atoms, such that percolation transition occurs only for $v \geq v_c(p)$.  
   
      For $p = 1$ one must include a non-trivial fraction $v_c(1)$ of similar bonds to achieve 
   a percolation transition. On a fully occupied lattice we have used again the bi-section method to obtain an
   accurate estimation of $v_c(1)$. Starting with two trial values of $v$ corresponding to the globally connected 
   and unconnected systems, the gap between them is reduced by successive halving of the interval. As before, the 
   values of $v_c(1,L)$ obtained this way have been extrapolated for $L \to \infty$ to obtain $v_c(1) = 0.0651(5)$. 
   It may be noted that the existence of a non-zero value of $v_c(1)$ is a numerical demonstration of the 
   absence of a percolation transition in the $n = 2$ colored percolation as well as in the $AB$ percolation on the 
   square lattice.

\begin{figure}[t]
\begin {center}
\includegraphics[width=6.0cm]{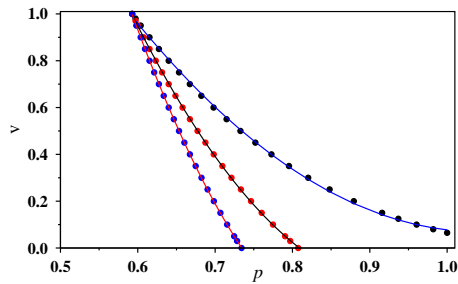}
\end {center}
\caption{Phase diagram of the density $v$ of similarly colored bonds and the site occupation probability $p$ for 
   $n$ = 2 (black), 3 (red), and 4 (blue) (arranged from right to left) with $q = 1/n$. The critical curve is fitted very closely by the Eqn. 
   \ref{EQN06} whose parameters are $c_1$ = 5.01, 8.28, and 11.47; $c_2$ = -9.48, -17.92, and -26.25; $c_3$ = 4.55, 
   9.50, and 14.47 for $n$ = 2, 3, and 4 respectively.
}
\label {FIG09}
\end{figure}

      In Fig. \ref{FIG08}(a), the order parameter $\Omega(v,L) = \langle s_{max}(v,L) \rangle / L^2$ has been plotted 
   against $v$ for three different sizes of the system. For $v = 0$, only the $AB$ bonds are present in the system and 
   the size of the largest cluster is minuscule, which is apparent by the very small value of the order parameter. On 
   increasing $v$ further, the order parameter grows monotonically and the sharpest rise occurs at a critical value 
   $v_c(1,L)$ leading to the occurrence of global connectivity. A finite-size scaling analysis is exhibited 
   in Fig. \ref{FIG08}(b) indicating a scaling form:
\begin{equation}
   \Omega(v,L)L^{\beta/\nu} \sim {\cal F}[(v - v_c(1))L^{1/\nu}]. 
\label{EQN05}
\end{equation} 
   Using $v_c = 0.0651$, the best data collapse is observed for $1/\nu = 0.745(5)$ and $\beta/\nu = 0.101(5)$, compared
   to the exact value of the correlation length exponent $1/\nu = 3/4$ and $\beta/\nu = 5/48 \approx 0.104$ for the 
   ordinary percolation in two dimensions \cite{Stauffer}. In addition, our estimates for the fractal dimension $d_f = 1.896(5)$ of
   the infinite incipient cluster \cite{Feder} and the exponent $\gamma = 2.388(5)$ of the second moment of the cluster size distribution
   at $v_c(1,L)$ yield values very much consistent with the exactly known exponents of $d_f = 91/48$ and $\gamma = 43/18$ 
   for the ordinary percolation which
   fulfill the scaling and hyperscaling relations in two dimensions: $\gamma/\nu$ + $2\beta/\nu = 2$ \cite{Stauffer,Ziff}.
   
      Repeating this method for many different values of occupation probability $p$ we have drawn the phase diagram
   in the $v - p$ plane in Fig. \ref{FIG09}. This plane is divided into two regions by the critical curve $v_c(p) = 
   {\cal G}(p)$ which separates the percolating region (above) from the non-percolating (below) region. Three different 
   critical curves are shown for $n$ = 2, 3, and 4. The dependence of the critical fraction $v_c(p)$ on $p$ for a 
   specific value of $n$ is obtained by the quadratic polynomial fit of the data as
   exhibited in Fig. \ref{FIG09}:
\begin{equation}
    v_c(p) = c_1 + c_2p + c_3p^2
\label{EQN06}
\end{equation} 
   The values of $c_1, c_2$, and $c_3$ are given in the Fig. \ref{FIG09} caption.
   
\begin{figure}[t]
\begin {center}
\includegraphics[width=7.0cm]{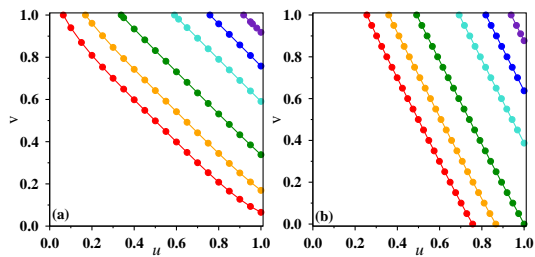}
\end {center}
\caption{Phase diagram of the density of similarly colored bonds $v$ and dissimilarly colored bonds $u$ for fixed 
   values of $p$ for $n = 2$ (a) and $n = 3$ (b), with $q = 1/n$ using $L = 1024$ on the square lattice. The values 
   of $p$ are 1.00 (red), 0.90 (orange), 0.80 (green), 0.70 (cyan), 0.65 (blue), and 0.61 (indigo) (arranged from 
   left to right). For every $p$, the region above the critical curve depicts the percolating phase.
}
\label {FIG10}
\end{figure}

\section {Generalized Colored Percolation with similar and dissimilar bonds}

      In this section we have generalized the model of colored percolation tuning the fractions of the bonds between 
   similar and dissimilar colored atoms using two independent parameters. Specifically, for the site occupation 
   probability $p > p_c$, the bonds between dissimilarly colored atoms are connected with probability $u$ and those 
   between similarly colored atoms are connected with probability $v$. Therefore, on the $u - v$ plane a critical 
   percolation curve represents the phase boundary between the percolating and the non-percolating phases.
   In Fig. \ref{FIG10}(a) we have shown for $n = 2$ and $q = 1/2$, a series of critical percolation curves for 
   different values of occupation probability $p$. Here, the density of connected bonds is given by $\text{Prob}_b(u,v) 
   = (u+v)/2$. The symmetry of this expression under the interchange of $u$ and $v$ is reflected by the mirror 
   symmetry of the curves in Fig. \ref{FIG10}(a) about the $v = u$ line. This can be generalized further for any
   value of $n$ as $\text{Prob}_b(u,v) = u+(v-u)/n$, and therefore, for $n > 2$ the critical curves are not symmetric 
   about the $u = v$ line any more. This has been exhibited in Fig. \ref{FIG10}(b) for $n = 3$ and $q = 1/3$.
   
\begin{figure}[t]
\begin {center}
\includegraphics[width=7.0cm]{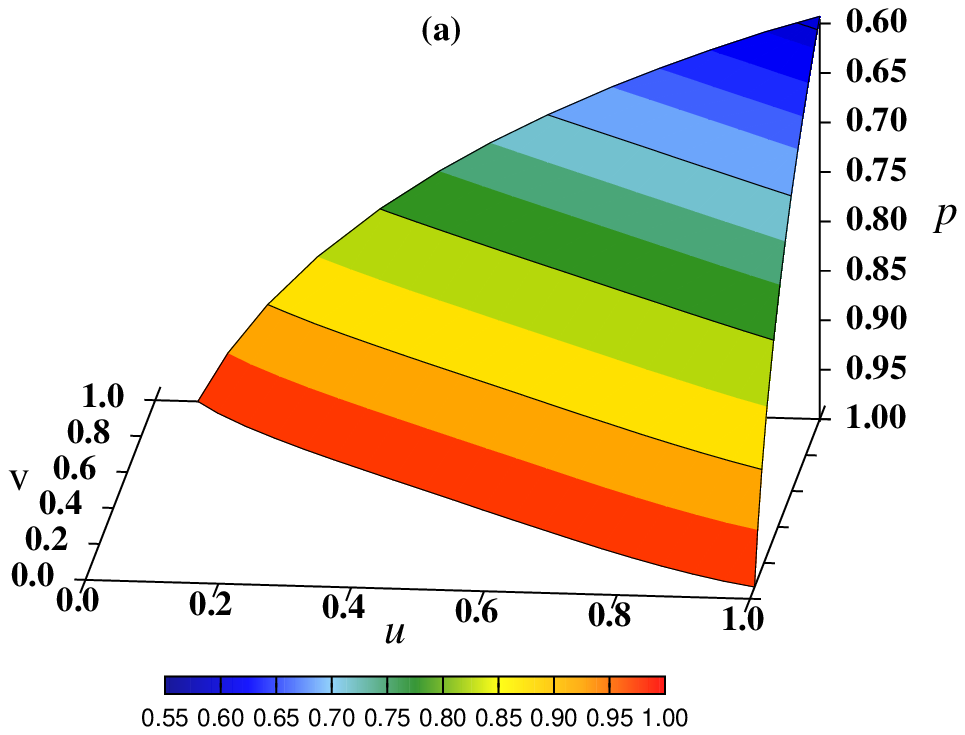}\\
\includegraphics[width=7.0cm]{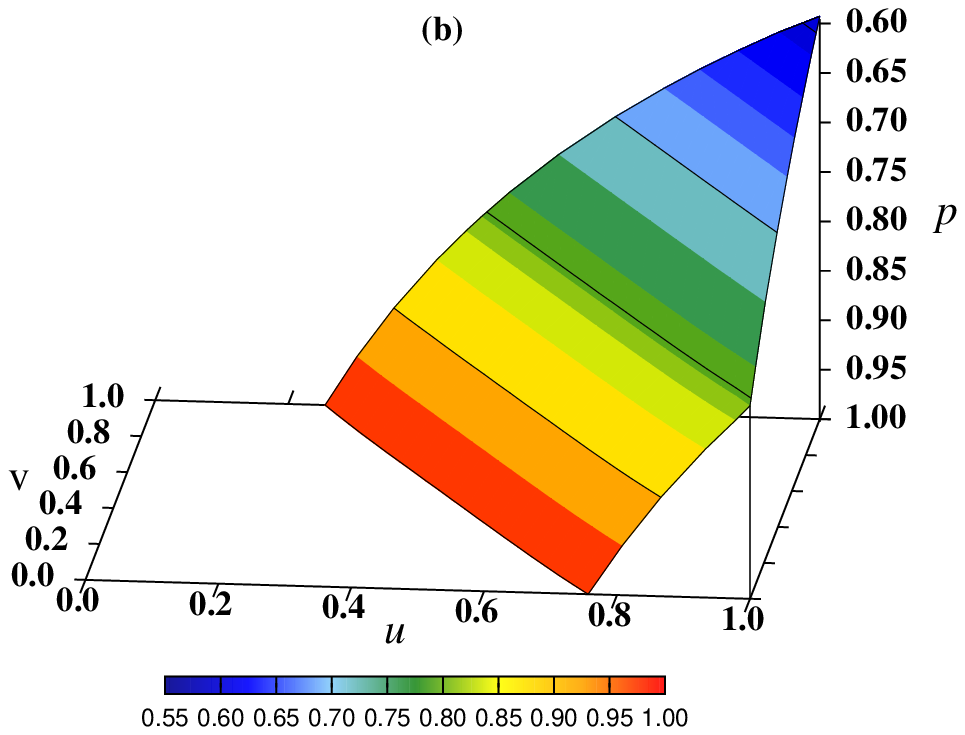}
\end {center}
\caption{$3$D phase diagram has been drawn in the $u - v - p$ plane, for $n = 2$ (a) and $n = 3$ (b), with $q = 1/n$ 
   using $L = 1024$ on the square lattice. The colored surface separates the percolating region from the non-percolating 
   region.
}
\label {FIG11}
\end{figure}

      A better visualization of the percolating and non-percolating phases has been exhibited by a three-dimensional 
   critical surface in the $(u - v - p)$ space. Fig. \ref{FIG11}(a) and (b) exhibit such plots for $n = 2$ and 3 
   respectively.  Any point within the space enclosed by the critical surface represents the percolating phase. The 
   intersections of these critical surfaces with the $u = 1$ plane have been shown in Fig. \ref{FIG09} for $n = 2$, 3
   and 4.
   
\section {Universality class of colored percolation}

      To confirm that the colored percolation belongs to the universality class of ordinary percolation,
   we have estimated a set of critical exponents, e. g., the fractal dimension of the largest cluster,
   the cluster size distribution exponent and the fractal dimension of the shortest paths right at 
   the percolation threshold of the unpreffered colored percolation. 

\begin{figure}[t]
\begin {center}
\includegraphics[width=6.0cm]{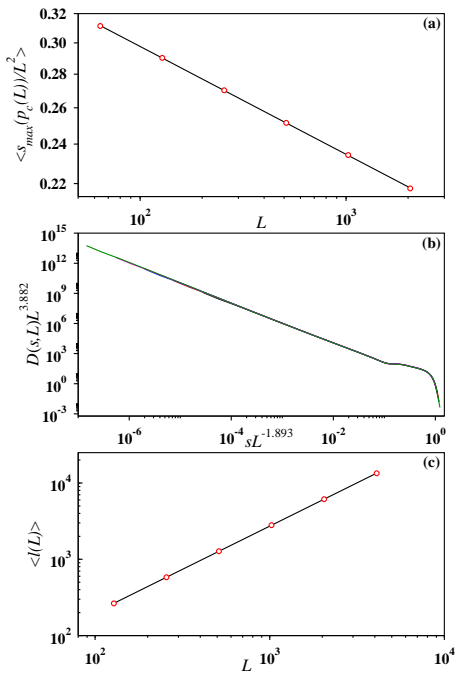}
\end {center}
\caption{Plots for $n$ = 3 and $q = 1/3$ at the percolation threshold of the colored percolation on the square lattice.
   (a) The average fractional size $\langle s_{max}(p_c(L)) \rangle / L^2$ of the largest cluster plotted against the 
   system size $L$ gives a value of the fractal dimension $d_f = 1.897(2)$.
   (b) The finite-size scaling analysis of the cluster size distribution $D(s,L)$ has been exhibited. Plot 
   of $D(s)L^{3.882}$ against $sL^{-1.893}$ shows an excellent data collapse, that yields
   $\tau = 2.051(5)$.
   (c) The average shortest path $\langle \ell(L) \rangle$ has been plotted against $L$ for $L$ = 128, 256, 
   512, 1024, 2048, and 4096. The fractal dimension of the shortest path is estimated from the slope as
   $d_{\ell} = 1.133(2)$.
}
\label {FIG12}
\end{figure}

      {\it Fractal dimension}: The average fractional size of the largest cluster at the percolation 
   threshold decreases with the system size $L$ as $\langle s_{max}(p_c(L)) \rangle /L^2 \sim L^{d_f - 2}$, 
   where $d_f$ is its fractal dimension \cite{Feder}. Our estimated values of $d_f = 1.897(2)$ for $n$ = 3 
   on square lattice (Fig. \ref{FIG12}(a)) and 1.895(2) for $n$ = 2 on triangular lattice are compared with 
   the fractal dimension 91/48 $\approx$ 1.8958 of the ordinary percolation in two dimensions.

      {\it Cluster Size Distribution}: The size $s$ of a percolation cluster being the number of occupied 
   sites in the cluster. Cluster sizes are measured for all clusters right at the percolation threshold,
   marked by the maximal jump of the largest cluster. The cluster size distribution $D(s)$ is measured by
   averaging over many different configurations. In Fig. \ref{FIG12}(b), the finite-size scaling of the 
   data for $D(s)$ has been shown for $n$ = 3 on the square lattice. An excellent collapse of the data
   confirms a power law variation: $D(s) \sim s^{-\tau}$. Using the best fitted values of the scaling exponents, 
   the cluster size distribution exponent $\tau$ has been estimated to be 2.051(5), compared to 187/91 
   $\approx 2.055$ for ordinary percolation in two dimensions \cite{Stauffer2,Isichenko}. 
   A very similar value of 2.051(5) has been found for $n = 2$ on the triangular lattice.

      {\it Shortest Path}: In general on a cluster, there exists multiple paths between an arbitrary pair of 
   sites. The smallest of these is called the shortest path and its length is measured by the number $\ell$
   of connected bonds on this path. Using the Burning algorithm \cite{Herrmann} the average lengths $\langle 
   \ell(n,L) \rangle$ of the system spanning shortest paths at the percolation threshold have been estimated 
   and is found to scale with the lattice size $L$ as $\langle \ell(n,L) \rangle \sim L^{d_{\ell}}$, with 
   $d_{\ell}$ = 1.133(2) for $n$ = 3 on the square lattice (Fig. \ref{FIG12}(c)) and 1.133(3) for $n$ = 2 on the 
   triangular lattice compared to $\approx 1.131$ in two dimension for the ordinary percolation \cite{Zhou,Schrenk}.

      The same set of critical exponents have been estimated for the preferential colored percolation 
   with $q$ = 0.60 and we have obtained very similar matching with the exponents of ordinary percolation.

\section {Summary}

      To summarize, we have introduced the model of `Colored Percolation' and have formulated its infinite number
   of versions in two dimensions. The sites of a regular lattice are occupied by atoms with probability $p$ and
   are colored randomly using one of the $n$ distinct colors with probability $q = 1/n$. A bond is said to be 
   connected if and only if its end atoms are of different colors. The global connectivity is then determined 
   through the connected bonds. It has been observed that the percolation threshold $p_c(n)$ approaches $p_c$ in 
   the limit of $n \to \infty$ as $1/n$.
   
       The preferential colored percolation has been defined when $m$ out of $n$ colors are selected with probability $q/m$
   each and rest of the colors are selected with probability $(1 - q)/(n - m)$. It has been observed that $p_c(q,m)$ depends 
   non-trivially on $q$ and has a minimum at $q_{min} = m/n$. The plot of $p_c(q,m)$ against $q$ is asymmetric for general
   value of $m$, but it becomes symmetric about $q = 1/2$ only when $m = n/2$. 
   
      This model is further generalized by adding similarly colored bonds of density $v$. It has been found that for
   each value of the site occupation probability $p$, there exists a non-trivial value of $v_c(p)$. The phase diagram
   in the $v - p$ plane has been drawn with $v = v_c(p)$ as the critical curve separating the percolating and non-percolating 
   regions. Such a phase diagram is better viewed in $3$D and drawn by tuning $v$ as well as $u$ of the fraction of 
   dissimilarly colored bonds and plotting the site occupation probability $p$ along the $z$-axis. Estimation of different 
   critical exponents lead us to conclude that all versions of the model of colored percolation belong to the percolation 
   universality class.
\section {Acknowledgments}

      We gratefully acknowledge R. M. Ziff and D. Dhar for their useful comments on this work.


\end {document}